\documentclass[useAMS,usenatbib]{mn2e}

\usepackage{graphicx}

\title[Ole R$\o$mer's method still on the stage: V994 Her]{Ole R$\o$mer's method still on the stage. The study of two bound eclipsing binaries in quintuple system V994 Her.}

\author[P.Zasche and R. Uhla\v{r}]{P.Zasche$^{1}$\thanks{E-mail:
zasche@sirrah.troja.mff.cuni.cz} and R. Uhla\v{r}$^{2}$\\
$^{1}$Astronomical Institute, Faculty of Mathematics and Physics, Charles University Prague, CZ-180 00 Praha 8,
V Hole\v{s}ovi\v{c}k\'ach 2,\\ Czech Republic\\
$^{2}$Private Observatory, Poho\v{r}\'{\i} 71, CZ-254 01 J\'{\i}lov\'e u Prahy, Czech Republic\\}
\begin{document}

\date{Accepted ... Received ...; in original form ...}

\pagerange{\pageref{firstpage}--\pageref{lastpage}} \pubyear{2012}

\maketitle

\label{firstpage}

\begin{abstract}
More than three hundred years ago, Ole R$\o$mer measured the speed of light only by observing the
periodic shifting of the observed eclipse arrival times of Jupiter's moons arising from the varying
Earth-Jupiter distance. The same method of measuring the periodic modulation of delays is still
used in astrophysics. The ideal laboratories for this effect are eclipsing binaries. The unique
system V994~Her consists of two eclipsing binaries orbiting each other. However, until now it was
not certain whether these are gravitationally bound and what their orbital period is. We show that
the system is in fact quintuple and the two eclipsing binaries are orbiting each other with period
about 6.3 years. This analysis was made only from studying the periodic modulation of the two
periods, when during the periastron passage one binary has an apparently shorter period, while the
other one longer, exactly as required by a theory. Additionally, it was found that both inner
eclipsing pairs orbit with slightly eccentric orbits undergoing a slow apsidal motions with a
period of the order of centuries.
\end{abstract}


\section{Introduction}

Eclipsing binaries are ideal astrophysical laboratories, and even more than a century of their
intensive photometric and spectroscopic monitoring, they still represent the best method to
determine the masses, radii and luminosities of stars. Thanks to modern ground- (and space-) based
telescopes we are able to discover these objects in other galaxies and to apply the same methods as
used in our Solar neighborhood. One very specific method is the analysis of their orbital periods.
Using the precise times of minima (centers of the eclipses of the components), we can determine if
the system's period is constant, accelerating, decelerating, or periodically alternating. If we
detect a periodic shifting of the times of minima, we determine that there is an additional
component in the system, which is orbiting around the barycenter with the eclipsing pair. If the
system is moderately inclined to the observer, the light from the eclipsing binary needs more and
less time to reach us, as it moves away and closer to the observer as it orbits the unseen
component. This method is in fact the same as used in the 17th century by Ole R$\o$mer when
measuring the finite speed of light using the eclipse times of Jupiter's moons, see e.g.
\cite{Romer}.

\section{The system V994 Her}

Dealing with the eclipsing binaries, we have a few advantages. First of all, there are currently
thousands of eclipsing binaries known. Moreover, the time baseline of observations for some
eclipsing binaries is more than a century long. Importantly, the observations are very easy to
obtain, even with small telescopes.

This is the case of one very interesting eclipsing system V994~Her (HD 170314, ADS 11373 AB; $V =
7.00$~mag). In 2008 it was discovered \citep{2008MNRAS.389.1630L} that V994~Her is the first (at
that time) system consisting of two eclipsing binaries. From one point on the sky we can see two
different sets of eclipses, one with the period of $P_A$ = 2.08 days, while the other one has a
period of $P_B$ = 1.42 days. The star was also observed with the Hipparcos satellite \citep{HIP},
however the strange eclipsing behavior was missed for about 15 years. The complete study of this
interesting system was made \citep{2008MNRAS.389.1630L} also on the basis of their new
spectroscopic observations, yielding a set of physical parameters of all four eclipsing components.
Their study showed that the system consists of two pairs: A (B8V + A0V), and B (A2V + A4V). All
components are well-detached and still located on the main sequence. Both orbits are slightly
eccentric.

\begin{table*}
 \centering
 \scalebox{0.96}{
 \begin{minipage}{180mm}
  \caption{List of currently known double eclipsing systems.}  \label{Tab1}
  \begin{tabular}{@{}c c c c c c c c@{}}
\hline
 System Name      & Other designation    & RA J2000.0  & DE J2000.0   &      V[mag]   & Period A [d] & Period B [d] & Type  \\
 \hline
OGLE LMC-ECL-16549 & OGLE LMC-SC3 179761 & 05 28 09.41 &$-$69 45 28.60&        18.216 &164.789640 & 0.818033 & EA + EW \\[2mm]
 CzeV343         & GSC 02405-01886       & 05 48 24.01 & +30 57 03.60 &        13.679 &  1.209373 & 0.806931 & EA + EA \\[2mm]
 TYC 3807-759-1  & GSC 03807-00759       & 09 30 10.75 & +53 38 59.80 &         9.538 &  1.305545 & 0.227715 & EA + EW \\[2mm]
 BV Dra          & HIP 74370             & 15 11 50.36 & +61 51 25.25 & $\,\,\,$8.040 &  0.350067 &          & EW + $\,$ \\
 BW Dra          & HIP 74368             & 15 11 50.09 & +61 51 41.16 & $\,\,\,$8.834 &           & 0.292165 &  $\,\,\,$ + EW \\[2mm]
 V994 Her        & GSC 02110-01170       & 18 27 45.89 & +24 41 50.66 & $\,\,\,$7.001 &  2.083269 & 1.420038 & EA + EA \\[2mm]
 KIC 4247791     & TYC 3124-1500-1       & 19 08 39.57 & +39 22 36.96 &        11.645 &  4.100871 & 4.049732 & EA + EA \\
 \hline
\end{tabular}
\end{minipage}}
\end{table*}

On the other hand, one important question arises, whether the two eclipsing components comprise one
gravitationally bounded system, or the system is only an optical binary. The mutual orbital period
of the two pairs can be very long and long-time monitoring is rather time consuming. The system
V994~Her also contains one more distant component observed interferometrically (\citealt{WDS}),
whose period was estimated of about a few thousand years. Therefore, the authors
\citep{2008MNRAS.389.1630L} speculated that the two eclipsing pairs could be identified with these
two visual components. For a brief review of quadruple systems with two eclipsing binaries see
\cite{2012A&A...544L...3C}. Currently we know only six such systems, BV + BW Dra, V994 Her, CzeV343
\citep{2012A&A...544L...3C}, KIC 4247791 \citep{2012A&A...541A.105L}, TYC 3807-759-1
\citep{2012arXiv1210.6765L}, and OGLE LMC-ECL-16549 \citep{2011AcA....61..103G}, see Table
\ref{Tab1}.

\section{Analysis}

Here we introduce an original approach of delay-time variations of both eclipsing pairs, showing
that the system is in fact quintuple and the two inner pairs are orbiting around each other with
much shorter period. We obtained many new observations of both pairs, as well as re-analyzed the
Hipparcos and ASAS \citep{2002AcA....52..397P} photometry. The individual times of minima are
presented in Table \ref{MINIMA}. These data were analyzed simultaneously in a combined approach of
fitting both orbits together, using a well-known method usually called "Light-time effect", or
"Light-travel-time effect" (hereafter LTTE), described elsewhere, e.g. \cite{Irwin1959} or
\cite{Mayer1990}.

This method has been used for dozens of binaries in the past, however, V994~Her is the first
eclipsing binary system, where the method can be applied to both binaries. The main advantage of
this approach is that both eclipsing pairs serve as strictly periodic "stellar lighthouses", whose
apparent period changes can be studied.

We used our new code for computing this combined approach of deriving both inner orbits (their
periods, and apsidal motion), together with the long orbit of mutual motion of the two pairs.
Altogether 15 parameters were fitted, using all available times-of-minima observations (for A and B
pairs: (25+36) published minima together with (18+17) our new unpublished data points).

In Figure 1 we plot the O-C diagrams for A and B pairs, showing their period changes (the y-axis)
with respect to the time (the x-axis). In these plots the positive y values indicate that the
detected signal occurs later, while the negative values indicate earlier detection than predicted
from strictly periodic linear behavior. As one can see, the rapid period changes near the
periastron passage are clearly visible for both pairs. Most of the parameters for the LTTE fits for
A and B are mostly the same. The exceptions are the omega angles (the argument of periastron,
$\omega_A$ = $\omega_B$ + 180$^\circ$), and the ($O-C$) amplitudes ($A_A$ and $A_B$, see below).
The long-term modulation arises from the apsidal motion of the inner orbits, because both are
slightly eccentric. For the final parameters of the fit see Table \ref{Tab_param}.

\begin{figure}
  \centering
  \includegraphics[width=85mm]{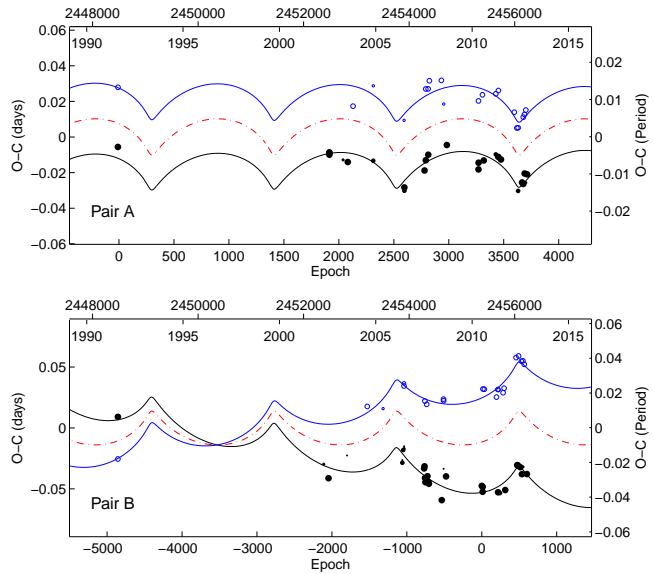}
  \caption{Plot of the O-C diagrams of both pairs. The dots stand for primary, while the open circles
  for the secondary minima, the bigger the symbol, the higher the weight. The red dash-dotted lines
  indicate the LTTE fit, while the black and blue curves represent the final fit (LTTE plus the
  apsidal motion).}
  \label{FigOC}
\end{figure}

\section{Results}

The main result of our analysis is the discovery that the two eclipsing pairs orbit around each
other and show also detectable period modulation together with a slow apsidal motion. The period of
the motion of apsides for pair A resulted in about (627 $\pm$ 439) years, while for pair B the
period is about (113 $\pm$ 10) years.

\begin{figure}
  \centering
  \includegraphics[width=85mm]{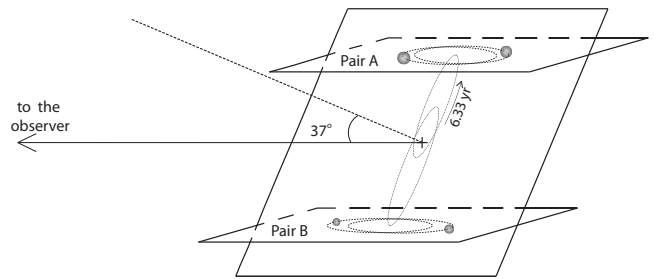}
  \caption{Schematic sketch of the V994 Her system is given (not to scale). The orbital planes of both A and B pairs are
  almost perpendicular to the celestial plane (i.e. edge-on to the observer, 84$^\circ$ and 86$^\circ$,
  respectively).}
  \label{Figure}
\end{figure}

\begin{table*}
 \centering
  \caption{V994~Her: The final parameters of the fits for A and B pairs.}  \label{Tab_param}
  \begin{tabular}{c c | c c | c c}
\hline
 & & \multicolumn{2}{c}{Pair A} & \multicolumn{2}{c}{Pair B} \\
 Parameter & Unit & Value & Error & Value  & Error \\
 \hline
 $JD_0$ &  HJD  & 2448501.1302 & 0.0110    & 2455375.4555 & 0.0062 \\
 $P$    &  Day  & 2.0832691    & 0.0000038 &  1.4200381   &  0.0000051 \\
 $e$    &       &   0.0311     &   0.0074  &   0.1258     &  0.0032 \\
 $\omega$& Deg  &    16.0      &    4.2    &   313.8      &  2.8   \\
 $\mathrm{d}\omega/\mathrm{d}t$ & Deg/Cycle & 0.0032 & 0.0013 & 0.0124 & 0.0010 \\
 $p_3$  & Year  &    6.33      &   0.56    &    6.33      &  0.56 \\
 $T_0$  &  HJD  & 2456067      &   199     &  2456067     & 199 \\
 $A$    & Day   &  0.0102      &   0.0042  &  0.0139      &  0.0057 \\
 $\omega_3$ & Deg &  256.2     &  24.1     &   76.2       &  24.1 \\
 $e_3$  &       &   0.747      &   0.182   &   0.747      &  0.182 \\
 \hline
\end{tabular}
\end{table*}

From the parameters of LTTE we determine the semi-major axis of the LTTE orbit, and using the
distance from the Hipparcos satellite \citep{HIP} of $\pi = (3.90 \pm 0.74)$~mas, we also derive
the angular separation of the two eclipsing binaries on the sky. This resulted in an angular
separation of $\Delta \alpha_{12} = (27.6 \pm 6.8)$ mas, which is well within the limits for modern
stellar interferometers, hence its discovery is expect soon. Moreover, this indicates that the
previously mentioned  star at $\sim$ 1 arcsec away is another star, not the eclipsing one as
suggested by \cite{2008MNRAS.389.1630L}. Membership of this distant component to the eclipsing
pairs was suggested via similar proper motions, see e.g. the WDS
catalogue\footnote{http://ad.usno.navy.mil/wds/}, \citealt{WDS}. Thus we have a quintuple star
system. We currently know only 20 quintuples \citep{2008MNRAS.389..869E}.

From the parameters of LTTE one can also calculate the mass function of the distant body
 $$f(m_3) =\frac{(m_3 \sin i)^3}{(m_1+m_2+m_3)^2} = \frac{1}{p_3^2} \left[ \frac{173.15 \, A}{\sqrt{1-e_3^2 \cos^2\omega}} \right]^3,$$
see e.g. \cite{Mayer1990}. Using the masses of both pairs as determined by
\cite{2008MNRAS.389.1630L}, we can also calculate the inclination between the eclipsing binary and
the LTTE orbit. If we label the inclination between the orbit of pair A and the LTTE orbit as
"$i_A$", and vice versa for B, then one can derive:
 $$\frac{m_A \cdot A_A}{\sin i_A} = \frac{m_B \cdot A_B}{\sin i_B}.$$
From this equation and the mass function found from the LTTE, we can directly determine the
inclination between the orbits. This results in $i_A= 37.1^\circ \pm 7.3^\circ$, while $i_B=
36.8^\circ \pm 6.8^\circ$. Evidently, the inclinations, as derived from both A and B pairs, are
comparable, and hence we know the absolute orientation of the orbit in space. This is the first
time that the mutual inclination between the orbits of the eclipsing pairs has been measured. Only
about twenty other systems with unambiguous mutual inclinations between the eclipsing and outer
orbits are known, see e.g. \cite{2011ApJ...728..111O}.

\section{Discussion and conclusions}

V994~Her is an interesting target for a future study. However, there still remain some open
questions. For example, these include a detection of the distant component in the spectra and
determination of its orbital period, etc. Additionally, some of the orbital elements of the 6.3-yr
orbit are still unknown - for instance the longitude of the ascending node $\Omega$. The long-term
evolution of outer and inner orbits should be studied over longer time scales. However, detecting
the mutual motion of the two eclipsing pairs is a unique discovery and we can still hope to find
similar configurations also in other multiple systems.

Surprisingly, for this analysis there was no need of spectroscopic observations for the radial
velocities of the long-period system, that would be rather complicated with current observing time
allocations on larger telescopes needed for such studies. It is noteworthy that all of our new
observations were carried out with 20-cm aperture or smaller telescopes by an amateur astronomer.
As clearly demonstrated by this study, scientifically valuable results can be secured with small
telescopes by amateur observers.

\section*{Acknowledgments}
The data used for the analysis are presented in the Appendix section in Table \ref{MINIMA}. We do
thank the "ASAS" team for making all of the observations easily public available. We also thank
Assoc.Prof. Marek Wolf for sending us his photometric data. Prof. Ed Guinan is also greatly
acknowledged for his referee comments, which significantly improved the manuscript. This work was
supported by the Czech Science Foundation grant no. P209/10/0715, by the research programme
MSM0021620860 of the Czech Ministry of Education, and by the grant UNCE 12 of the Charles
University in Prague. This research has made use of the Washington Double Star Catalog maintained
at the U.S. Naval Observatory, the SIMBAD database, operated at CDS, Strasbourg, France, and of
NASA's Astrophysics Data System Bibliographic Services.

\newpage

\appendix

\section[]{}

\label{lastpage}

\begin{table}
 \centering
 \tiny
 \begin{minipage}{90mm}
  \caption{Heliocentric minima of V994 Her.} \label{MINIMA}
  \begin{tabular}{@{}l l l l l l@{}}
\hline
HJD - 2400000 & Error & Pair & Type & Filter & Observer/Reference\\
 \hline
 48480.29371 & 0.00125 & A  & p  &  Hp   &    Hipparcos - this paper                  \\
 48481.36885 & 0.00169 & A  & s  &  Hp   &    Hipparcos - this paper                  \\
 52488.4984  & 0.0006  & A  & p  &  R    &    \cite{2002IBVS.5313....1B}              \\
 52488.4994  & 0.0001  & A  & p  &  V    &    \cite{2002IBVS.5313....1B}              \\
 52488.4997  & 0.0003  & A  & p  &  B    &    \cite{2002IBVS.5313....1B}              \\
 52748.9041  & 0.0010  & A  & p  &  V    &    San Pedro Martir, M.Wolf - this paper   \\
 52836.4002  & 0.0002  & A  & p  &  R    &    \cite{2004IBVS.5579....1B}              \\
 52937.4701  & 0.0001  & A  & s  &  R    &    \cite{2006IBVS.5684....1B}              \\
 53319.71928 & 0.00239 & A  & p  &  V    &    ASAS - this paper                       \\
 53320.80295 & 0.00542 & A  & s  &  V    &    ASAS - this paper                       \\
 53904.09882 & 0.00169 & A  & s  &  V    &    \cite{2008MNRAS.389.1630L}              \\
 53909.26748 & 0.00089 & A  & p  &  B    &    \cite{2008MNRAS.389.1630L}              \\
 53909.26932 & 0.00072 & A  & p  &  V    &    \cite{2008MNRAS.389.1630L}              \\
 54290.517   & 0.0010  & A  & p  &  V    &    \cite{2011IBVS.5979....1B}              \\
 54314.5204  & 0.0008  & A  & s  &  V    &    \cite{2011IBVS.5979....1B}              \\
 54315.522   & 0.0010  & A  & p  &  V    &    \cite{2011IBVS.5979....1B}              \\
 54360.3524  & 0.0004  & A  & s  &  V    &    \cite{2011IBVS.5979....1B}              \\
 54361.3570  & 0.0002  & A  & p  &  V    &    \cite{2011IBVS.5979....1B}              \\
 54383.2729  & 0.0003  & A  & s  &  V    &    \cite{2011IBVS.5979....1B}              \\
 54610.3494  & 0.0008  & A  & s  &  BVR  &    \cite{2011IBVS.5979....1B}              \\
 54654.08476 & 0.00256 & A  & s  &  V    &    ASAS - this paper                       \\
 54713.4349  & 0.0014  & A  & p  &  BVR  &    \cite{2011IBVS.5979....1B}              \\
 55313.4026  & 0.00068 & A  & p  &  R    &    \cite{2011IBVS.6007....1Z}              \\
 55314.4827  & 0.00046 & A  & s  &  I    &    \cite{2011IBVS.6007....1Z}              \\
 55315.4896  & 0.00042 & A  & p  &  I    &    \cite{2011IBVS.6007....1Z}              \\
 55389.48381 & 0.00225 & A  & s  &  I    &    \cite{2011IBVS.6007....1Z}              \\
 55413.40450 & 0.0010  & A  & p  &  VRI  &    \cite{2011OEJV..137....1B}              \\
 55641.55985 & 0.00012 & A  & s  &  R    &    \cite{2011IBVS.6007....1Z}              \\
 55642.56748 & 0.00042 & A  & p  &  R    &    \cite{2011IBVS.6007....1Z}              \\
 55688.39802 & 0.00029 & A  & p  &  R    &    \cite{2011IBVS.6007....1Z}              \\
 55691.56016 & 0.00011 & A  & s  &  I    &    \cite{2011IBVS.6007....1Z}              \\
 55740.47826 & 0.00135 & A  & p  &  BVRI &    J.Trnka - var.astro.cz                  \\
 55993.62195 & 0.00007 & A  & s  &  I    &    R.Uhla\v{r} - this paper                \\
 56041.52826 & 0.00101 & A  & s  &  I    &    R.Uhla\v{r} - this paper                \\
 56064.44435 & 0.00049 & A  & s  &  C    &    R.Uhla\v{r} - this paper                \\
 56065.45050 & 0.00159 & A  & p  &  R    &    R.Uhla\v{r} - this paper                \\
 56138.36968 & 0.00059 & A  & p  &  I    &    R.Uhla\v{r} - this paper                \\
 56162.36395 & 0.00040 & A  & s  &  R    &    R.Uhla\v{r} - this paper                \\
 56163.36806 & 0.00026 & A  & p  &  R    &    R.Uhla\v{r} - this paper                \\
 56187.36470 & 0.00076 & A  & s  &  IC   &    R.Uhla\v{r} - this paper                \\
 56188.37316 & 0.00054 & A  & p  &  I    &    R.Uhla\v{r} - this paper                \\
 56210.28311 & 0.00180 & A  & s & C      &   R.Uhla\v{r} - this paper                 \\
 56211.28403 & 0.00107 & A  & p & C      &   R.Uhla\v{r} - this paper                 \\
 56234.20460 & 0.00183 & A  & p & R      &   R.Uhla\v{r} - this paper                 \\ \hline
 48479.75947 & 0.00305 & B  & p  & Hp    &   Hipparcos - this paper                   \\
 48480.43493 & 0.00263 & B  & s  & Hp    &   Hipparcos - this paper                   \\
 52471.4363  & 0.0004  & B  & p  & UBV   &   \cite{2003IBVS.5462....1A}               \\
 52381.9854  & 0.0010  & B  & p  & V     &   San Pedro Martir, M.Wolf - this paper    \\
 52822.20440 & 0.00185 & B  & p  & V     &   ASAS - this paper                        \\
 53206.3650  & 0.002   & B  & s  & VR    &   \cite{2007IBVS.5753....1B}               \\
 53504.57119 & 0.00454 & B  & s  & V     &   ASAS - this paper                        \\
 53870.18880 & 0.00255 & B  & p  & B     &   \cite{2008MNRAS.389.1630L}               \\
 53870.18658 & 0.00154 & B  & p  & V     &   \cite{2008MNRAS.389.1630L}               \\
 53887.23816 & 0.00216 & B  & p  & B     &   \cite{2008MNRAS.389.1630L}               \\
 53887.23754 & 0.00107 & B  & p  & V     &   \cite{2008MNRAS.389.1630L}               \\
 53902.20258 & 0.00097 & B  & s  & B     &   \cite{2008MNRAS.389.1630L}               \\
 53902.20047 & 0.00048 & B  & s  & V     &   \cite{2008MNRAS.389.1630L}               \\
 53904.28105 & 0.00170 & B  & p  & B     &   \cite{2008MNRAS.389.1630L}               \\
 53904.27915 & 0.00265 & B  & p  & V     &   \cite{2008MNRAS.389.1630L}               \\
 54283.4131  & 0.0001  & B  & p  & V     &   \cite{2011IBVS.5979....1B}               \\
 54290.515   & 0.003   & B  & p  & V     &   \cite{2011IBVS.5979....1B}               \\
 54298.3787  & 0.0002  & B  & s  & V     &   \cite{2011IBVS.5979....1B}               \\
 54300.4457  & 0.0004  & B  & p  & V     &   \cite{2011IBVS.5979....1B}               \\
 54307.5424  & 0.0005  & B  & p  & V     &   \cite{2011IBVS.5979....1B}               \\
 54332.457   & 0.001   & B  & s  & V     &   \cite{2011IBVS.5979....1B}               \\
 54334.523   &         & B  & p  & V     &   \cite{2011IBVS.5979....1B}               \\
 54347.3084  & 0.0003  & B  & p  & V     &   \cite{2011IBVS.5979....1B}               \\
 54364.3439  & 0.0003  & B  & p  & V     &   \cite{2011IBVS.5979....1B}               \\
 54374.2829  & 0.0007  & B  & p  & V     &   \cite{2011IBVS.5979....1B}               \\
 54618.5160  & 0.0013  & B  & p  & BVR   &   \cite{2011IBVS.5979....1B}               \\
 54650.5498  & 0.0030  & B  & s  & BVR   &   \cite{2011IBVS.5979....1B}               \\
 54652.62262 & 0.00681 & B  & p  & V     &   ASAS - this paper                        \\
 54653.3886  & 0.0013  & B  & s  & BVR   &   \cite{2011IBVS.5979....1B}               \\
 54699.4777  & 0.0014  & B  & p  & BVR   &   \cite{2011IBVS.5979....1B}               \\
 55375.40806 & 0.00037 & B  & p  & I     &   \cite{2011IBVS.6007....1Z}               \\
 55392.44365 & 0.0009  & B  & p  & RI    &   \cite{2011OEJV..137....1B}               \\
 55392.44779 & 0.00055 & B  & p  & I     &   \cite{2011IBVS.6007....1Z}               \\
 55397.49810 & 0.00032 & B  & s  & I     &   \cite{2011IBVS.6007....1Z}               \\
 55424.47871 & 0.00079 & B  & s  & R     &   \cite{2011IBVS.6007....1Z}               \\
 55654.51845 & 0.00097 & B  & s  & C     &   \cite{2011IBVS.6007....1Z}               \\
 55681.50533 & 0.00036 & B  & s  & R     &   \cite{2011IBVS.6007....1Z}               \\
 55683.55108 & 0.00024 & B  & p  & R     &   \cite{2011IBVS.6007....1Z}               \\
 55691.44530 & 0.00013 & B  & s  & I     &   \cite{2011IBVS.6007....1Z}               \\
 55779.48535 & 0.00125 & B  & s  & R     &   \cite{2011IBVS.6007....1Z}               \\
 55799.36956 & 0.00033 & B  & s  & I     &   \cite{2011IBVS.6007....1Z}               \\
 55821.29667 & 0.00095 & B  & p  & RV    &   L.\v{S}melcer - var.astro.cz             \\
 56026.60079 & 0.00063 & B  & s  & C     &   R.Uhla\v{r} - this paper                 \\
 56048.52317 & 0.00028 & B  & p  & R     &   R.Uhla\v{r} - this paper                 \\
 56058.46317 & 0.00020 & B  & p  & I     &   R.Uhla\v{r} - this paper                 \\
 56073.46347 & 0.00029 & B  & s  & R     &   R.Uhla\v{r} - this paper                 \\
 56102.48296 & 0.00058 & B  & p  & R     &   R.Uhla\v{r} - this paper                 \\
 56127.42060 & 0.00062 & B  & s  & R     &   R.Uhla\v{r} - this paper                 \\
 56136.55813 & 0.00106 & B  & p  & R     &   R.Uhla\v{r} - this paper                 \\
 56154.40165 & 0.00039 & B  & s  & C     &   R.Uhla\v{r} - this paper                 \\
 56156.44467 & 0.00159 & B  & p  & C     &   R.Uhla\v{r} - this paper                 \\
 56181.37945 & 0.00018 & B  & s  & R     &   R.Uhla\v{r} - this paper                 \\
 56230.28075 & 0.00063 & B  & p & C      &   R.Uhla\v{r} - this paper                 \\
  \hline
 \end{tabular}
\end{minipage}
\end{table}

\end{document}